\documentclass[prd,twocolumn,showpacs,amsmath,amssymb,superscriptaddress,preprintnumbers,nofootinbib]{revtex4}

\usepackage{amssymb}
\usepackage{amsmath}
\usepackage{graphicx}

\newcommand{\D}{\hat{D}}

\begin{document}

\title{Regular nonminimal magnetic
black holes \\
in spacetimes with a cosmological constant}

\author{Alexander B. Balakin}
\email{Alexander.Balakin@kpfu.ru} \affiliation{Department of
General Relativity and Gravitation, Institute of Physics, Kazan
Federal University, Kremlevskaya str.~18, Kazan 420008, Russia}

\author{Jos\'e P. S. Lemos}
\email{joselemos@ist.utl.pt} \affiliation{ Departamento de F\'{\i}sica,
Centro
Multidisciplinar de Astrof\'{\i}sica-CENTRA,
Instituto Superior T\'ecnico-IST, Universidade  de
Lisboa-UL, \\ Avenida Rovisco Pais 1, 1049-001 Lisboa, Portugal}

\author{Alexei E. Zayats}
\email{Alexei.Zayats@kpfu.ru} \affiliation{Department of General
Relativity and Gravitation, Institute of Physics, Kazan Federal
University, Kremlevskaya str.~18, Kazan 420008, Russia}

\begin{abstract}
We consider new regular exact spherically symmetric
solutions of a nonminimal
Einstein--Yang-Mills theory with a
cosmological constant and a gauge field of magnetic
Wu-Yang type. The most interesting solutions found are
black holes
with metric and curvature
invariants that are
regular everywhere, i.e., regular black holes.
We
set up a classification of the solutions
according to the number and type of horizons.
The structure of these regular black holes is
characterized by four specific features:
a small cavity in the neighborhood of the center,
a repulsion
barrier off the small cavity, a distant
equilibrium point, in which the metric function has a minimum,
and a
region of Newtonian attraction.
Depending on the sign and value
of the cosmological constant
the solutions are asymptotically de Sitter (dS),
asymptotically flat, or asymptotically anti-de Sitter (AdS).
\end{abstract}


\maketitle

\section{Introduction}

The interest in nonminimal theories,
i.e., theories that couple the gravitational field
to other fields using cross terms containing the curvature tensor,
started to rise as alternative theories of gravity
a long time ago.

The nonminimal field theories are based on
five classes, divided accordingly to the types of fields that
couple nonminimally to gravitation.
The first class of nonminimal models deals with the
coupling of scalar fields with the spacetime curvature.
They arose first within the Scherrer--Jordan--Thiry--Brans-Dicke
theory where a scalar field couples nonminimally
to the Ricci scalar, and the different authors
\cite{scherrer,jordan,thiry,bd}
had different motivations to set up
such a theory, see \cite{goenner} for a historical perspective.
In addition a scalar field conformally coupled to gravitation
was suggested in \cite{ccj}, see
\cite{NMscal1} for a review on the subject.
The second class is based on the modeling of nonminimal interactions
of the electromagnetic field with curvature, usually called nonminimal
Einstein-Maxwell model, see, e.g., \cite{NM1,NM2,NM3,BL2005}.  The
third class comprises Einstein--Yang-Mills models with $SU(n)$ symmetry
\cite{Horn81,MH2}.  The fourth class embodies
Einstein--Yang-Mills--Higgs models \cite{BaDeZa07,BaDeZa08}.  And
the fifth class covers models in which an axion pseudoscalar field
appears, for instance coupling to the electromagnetic and
gravitational fields in what may be called nonminimal
Einstein-Maxwell-axion models \cite{BNi2010}.

Within these nonminimal theories exact solutions for
electric stars \cite{HornPRD,MHS}, magnetic stars \cite{HornJMP},
wormholes \cite{BSZ07,BLZ10,DF2014}, electric black holes
\cite{BBL,BZ2015}
(where star-like solutions can also be found),
and regular magnetic  black holes
\cite{BaZaPLB}, with a
$SU(2)$ Wu-Yang ansatz
(see \cite{wuyang} for the original ansatz and work,
and \cite{shnir} for a full study of its properties) have been constructed.
Various aspects of star and black hole physics were discussed in these
papers in all five classes of the mentioned
nonminimally coupled theories.

Regular black holes are black holes without
spacetime curvature singularities and
have been constructed
with general relativity
coupled to some form of matter or within
alternative and extended theories of gravity.
For a sample of regular black holes see
\cite{b68,dy92,bor,bij,ab00,mat,horv,mat2,bdm,LZ11,bbs,yosh,azreg,bala,good,tosh,ma}.

In this vein, we want to further investigate nonminimal regular
magnetic black holes in spacetimes with a cosmological constant,
extended thus the work where no cosmological constant is considered
\cite{BaZaPLB}.
We discuss new examples of such exact solutions
of a $\Lambda$ nonminimal $SU(2)$-symmetric theory of a gauge field of
magnetic type satisfying the Wu-Yang ansatz.
Depending on the sign and value
of the cosmological constant
the solutions are asymptotically de Sitter (dS),
asymptotically flat, or asymptotically anti-de Sitter (AdS).
The new features
appearing with the $\Lambda$ term in the structure of the black holes
with a regular center are certainly of interest.  If the dark energy
is indeed a cosmological constant term and if a fundamental theory is
nonminimal then our study displays the role of the dark energy on the
regularization of the internal structure of black holes.

The paper is organized as follows. In Sec.~\ref{sec2}
we set up the general formalism
for a nonminimal Einstein--Yang-Mills theory.
In Sec.~\ref{specific} we present a
specific choice for the nonminimal parameters
and display the important reduced master equation.
In Sec.~\ref{generic} we investigate
the properties and the horizons
of the exact solution.
A thorough analysis is made
of the solutions, in particular of the regular
magnetic black hole solutions, for positive
cosmological constant in Sec.~\ref{g0}, for zero
cosmological constant in Sec.~\ref{eq0}, and for negative
cosmological constant in Sec.~\ref{l0}.
In Sec.~\ref{disc} we conclude.

\section{Nonminimal Einstein--Yang-Mills theory: formalism
and equations for static spherically symmetric
Wu-Yang-type magnetic objects}\label{sec2}

\subsection{General formalism and master equations}

We study a nonminimal Einstein--Yang-Mills
theory with $SU(2)$ symmetry and a
Wu-Yang-type ansatz as proposed in \cite{BaZaPLB}.
This theory is a generalization of the
nonminimal Einstein-Maxwell theory
with $U(1)$ symmetry studied in
\cite{BL2005}. The basic elements of this
Einstein--Yang-Mills
theory are given now.

The  nonminimal Einstein--Yang-Mills theory can be
described in terms of the action functional
\begin{align}
S_{{\rm NMEYM}} &= \int d^4 x \sqrt{-g}\
\left\{\frac{R+2\Lambda}{8\pi}+\frac{1}{2}F^{(a)}_{ik}
F^{ik(a)} \right.\nonumber\\
&\left.{}+\frac{1}{2}{\cal R}^{ikmn}F^{(a)}_{ik} F^{(a)}_{mn}
\right\}\,.\label{act}
\end{align}
Here $g = {\rm det}(g_{ik})$ is the determinant of the metric tensor
$g_{ik}$, $R$ is the Ricci scalar, $\Lambda$ is the cosmological
constant, and
$8\pi$ is the coupling constant (we are putting
the speed of light and the gravitational constant equal
to one, $c=1$, $G=1$,
so that $\frac{8\pi G}{c^4}=8 \pi$). Latin indices
without parentheses run from 0 to 3.
Group indices, $(a)$, run from 1 to 3, and when repeated
should be summed with a Kronecker delta metric.
The nonminimal
susceptibility tensor ${\cal R}^{ikmn}$ is defined as
\begin{align}
{\cal R}^{ikmn} &\equiv \frac{q_1}{2}R\,(g^{im}g^{kn} -
g^{in}g^{km})\nonumber\\ &{}+ \frac{q_2}{2}(R^{im}g^{kn} - R^{in}g^{km} +
R^{kn}g^{im} -R^{km}g^{in})\nonumber \\ &{}+ q_3 R^{ikmn}\,, \label{sus}
\end{align}
where $R^{ik}$ and $R^{ikmn}$ are the Ricci and Riemann tensors,
respectively, and $q_1$, $q_2$, $q_3$ are the phenomeno\-logi\-cal
parameters describing the nonminimal coupling of the Yang-Mills
field with the gravitational field.
The
$SU(2)$ Yang-Mills field is described by a triplet of
vector potentials $A^{(a)}_m$, where the group index $(a)$ runs
over three values. The Yang-Mills field
components $F^{(a)}_{mn}$ are
connected to the potentials $A^{(a)}_i$ by the
formulas
\begin{equation}
F^{(a)}_{mn} = \nabla_m A^{(a)}_n - \nabla_n A^{(a)}_m +
f^{(a)}_{\cdot (b)(c)} A^{(b)}_m A^{(c)}_n \,. \label{Fmn}
\end{equation}
Here $\nabla _m$ is a  covariant spacetime derivative, and
the
symbols $f^{(a)}_{\cdot (b)(c)}$ denote the real structure
constants of the gauge group $SU(2)$.

The variation of the action
(\ref{act}) with respect to the Yang-Mills potential $A_i^{(a)}$
yields
\begin{equation}
{\D}_k H^{(a)ik} =0 \,,
\label{YMeq0}
\end{equation}
with
\begin{equation}
H^{(a)ik} = F^{(a)ik} + {\cal
R}^{ikmn} F^{(a)}_{mn} \,. \label{YMeq}
\end{equation}
The tensor $H^{(a)ik}$ is the non-Abelian analog of the excitation
tensor in electrodynamics. This analogy allows us
to consider ${\cal R}^{ikmn}$ as a susceptibility tensor. The
gauge covariant derivative $\D_m$ of an arbitrary tensor $Q^{(a)
\cdot \cdot \cdot}_{\cdot \cdot \cdot (d)}$ is defined as follows
\begin{align}
\D_m Q^{(a) \cdot \cdot \cdot}_{\cdot \cdot \cdot (d)} &\equiv
\nabla_m Q^{(a) \cdot \cdot \cdot}_{\cdot \cdot \cdot (d)} +
f^{(a)}_{\cdot (b)(c)} A^{(b)}_m Q^{(c) \cdot \cdot
\cdot}_{\cdot \cdot \cdot (d)}+\dots\nonumber\\
&{} - f^{(c)}_{\cdot (b)(d)}
A^{(b)}_m Q^{(a) \cdot \cdot \cdot}_{\cdot \cdot \cdot (c)} -\dots
\,. \label{DQ2}
\end{align}
The variation of the action functional Eq.~(\ref{act})
with
respect to the metric $g_{ik}$ yields
\begin{equation}
R_{ik}-\frac{1}{2}Rg_{ik}= \Lambda g_{ik}+8\pi T^{\rm
(eff)}_{ik}\,.\label{EinMaster}
\end{equation}
The effective stress-energy tensor $T^{({\rm eff})}_{ik}$
appearing in Eq.~(\ref{EinMaster})
 can be
divided into four parts,
\begin{equation}
T^{\rm (eff)}_{ik} =  T^{(YM)}_{ik} + q_1 T^{(I)}_{ik} + q_2
T^{(II)}_{ik} + q_3 T^{(III)}_{ik} \,. \label{Tdecomp}
\end{equation}
The first term
\begin{equation}
T^{(YM)}_{ik} \equiv \frac{1}{4} g_{ik} F^{(a)}_{mn}F^{mn(a)} -
F^{(a)}_{in}F_{k}^{\ n(a)}  \label{TYM}
\end{equation}
is the stress-energy tensor of the pure Yang-Mills field. The
definitions of the other three tensors are related to the
corresponding coupling constants $q_1$, $q_2$, $q_3$.
So
\begin{align}
T^{(I)}_{ik} &= R\,T^{(YM)}_{ik} -  \frac{1}{2} R_{ik}
F^{(a)}_{mn}F^{mn(a)} \nonumber\\ &{}+\frac{1}{2}
\left[ {\D}_{i} {\D}_{k} -
g_{ik} {\D}^l {\D}_l \right] \left[F^{(a)}_{mn}F^{mn(a)} \right]
\,, \label{TI}
\end{align}
\begin{align}
T^{(II)}_{ik} &=\frac{1}{2}{\D}_l \left[
{\D}_i \left( F^{(a)}_{kn}F^{ln(a)} \right) {+} {\D}_k
\left(F^{(a)}_{in}F^{ln(a)} \right) \right] \nonumber\\
&{}-\frac{1}{2}g_{ik}\biggl[{\D}_{m}
{\D}_{l}\left(F^{mn(a)}F^{l(a)}_{\ n}\right)-R_{lm}F^{mn (a)}
F^{l(a)}_{\ n} \biggr]\nonumber \\
&{}- F^{ln(a)} \left(R_{il}F^{(a)}_{kn} +
R_{kl}F^{(a)}_{in}\right) \nonumber\\
&{}- R^{mn}F^{(a)}_{im} F_{kn}^{(a)} - \frac{1}{2} {\D}^m{\D}_m
\left(F^{(a)}_{in} F_{k}^{ \ n(a)}\right), \label{TII}
\end{align}
\begin{align}
T^{(III)}_{ik} &= \frac{1}{4}g_{ik}
R^{mnls}F^{(a)}_{mn}F_{ls}^{(a)}\nonumber\\
&{}- \frac{3}{4} F^{ls(a)}
\left(F_{i}^{\ n(a)} R_{knls} + F_{k}^{\
n(a)}R_{inls}\right)\nonumber \\
&{}- \frac{1}{2}{\D}_{m} {\D}_{n} \left[ F_{i}^{ \
n (a)}F_{k}^{ \ m(a)} + F_{k}^{ \ n(a)} F_{i}^{ \ m(a)} \right]
\,. \label{TIII}
\end{align}
Now we consider the master equations for static
spherically symmetric spacetimes and objects.

\subsection{Spherical symmetric ansatz,
nonminimal Wu-Yang-type magnetic
ansatz and the equations}

Let us consider a static spherically symmetric spacetime with
metric
\begin{equation}
ds^2= N dt^2-\frac{dr^2}{N}-r^2 \left( d\theta^2 + \sin^2\theta
d\varphi^2 \right) \,.\label{metrica}
\end{equation}
Here $(t,r,\theta,\varphi)$ are spacetime spherical
coordinates and
$N$ is a function depending on the radial variable $r$ only.

The gauge field is considered to be characterized by the
Wu-Yang ansatz
(see, e.g., \cite{BaZaPLB}). We use a
position dependent basis ${\bf t}_{(r)}$, ${\bf
t}_{(\theta)}$, ${\bf t}_{(\varphi)}$ for the generators
of the $SU(2)$ group, instead of the
standard basis ${\bf t}_{(1)}$,
${\bf t}_{(2)}$, ${\bf t}_{(3)}$. The
generators ${\bf t}_{(r)}$, ${\bf
t}_{(\theta)}$, ${\bf t}_{(\varphi)}$
are defined in terms of ${\bf t}_{(1)}$,
${\bf t}_{(2)}$, ${\bf t}_{(3)}$ as
\begin{gather}
{\bf t}_{(r)}=\cos{\nu\varphi} \ \sin{\theta}\;{\bf
t}_{(1)}+\sin{\nu\varphi} \ \sin{\theta}\;{\bf
t}_{(2)}+\cos{\theta}\;{\bf t}_{(3)} \,,\nonumber\\
{\bf t}_{(\theta)}=\cos{\nu\varphi} \ \cos{\theta}\;{\bf
t}_{(1)}+\sin{\nu\varphi} \ \cos{\theta}\;{\bf
t}_{(2)}-\sin{\theta}\;{\bf t}_{(3)} \,,\nonumber\\
{\bf t}_{(\varphi)}= -  \sin{\nu \varphi}\;{\bf
t}_{(1)}+\cos{\nu\varphi} \;{\bf t}_{(2)} \,, \label{deS5}
\end{gather}
and satisfy the commutation relations
\begin{equation}
\left[{\bf t}_{r},{\bf t}_{\theta}\right]={\bf t}_{\varphi}
\,,\quad \left[{\bf t}_{\theta} \,, {\bf
t}_{\varphi}\right]={\bf t}_{r} \,, \quad \left[{\bf
t}_{\varphi},{\bf t}_{r}\right]={\bf t}_{\theta}\,.\label{deS6}
\end{equation}
The Wu-Yang-type ansatz is of the form (see \cite{BaZaPLB} for
details)
\begin{gather}
A^{(a)}_{0}=0 \,, \quad A^{(a)}_{r}=0 \,,\nonumber\\
A^{(a)}_{\theta}=
\delta^{(a)}_{(\varphi)}  \,, \quad
A^{(a)}_{\varphi}=-\nu\sin{\theta} \ \delta^{(a)}_{(\theta)}\,.\label{1}
\end{gather}
The magnetic parameter $\nu$ is a nonvanishing integer. The field strength
tensor has only one nonvanishing component,
\begin{equation}
F^{(r)}_{\theta\varphi}= - A^{(\varphi)}_{\theta}
A^{(\theta)}_{\varphi}=\nu \sin{\theta}\,.\label{2}
\end{equation}
Clearly, it is a magnetic-type solution and it does not
depend on the parameters $\Lambda$, $q_1$, $q_2$ and $q_3$.

Then, the
nonminimal gravitational field equations reduced to spherical
symmetry have the form
\begin{gather}
\frac{1-N}{r^2}-\frac{N'}{r} - \Lambda = \frac{8\pi
\nu^2}{r^4}\left[\frac{1}{2}-
q_1\frac{N'}{r}\right.\nonumber\\ \left.{}+(13q_1+4q_2+q_3)\frac{N}{r^2}-
\frac{q_1+q_2+q_3}{r^2}\right], \label{G1}
\end{gather}
\begin{gather}
\frac{1-N}{r^2}-\frac{N'}{r}- \Lambda = \frac{8\pi
\nu^2}{r^4}\left[\frac{1}{2}- q_1
\frac{N'}{r}\right.\nonumber\\ \left. {}-(7q_1+4q_2+q_3)\frac{N}{r^2}-
\frac{q_1+q_2+q_3}{r^2}\right] , \label{G2}
\end{gather}
\begin{gather}
\frac{1}{r}N'+\frac{1}{2} N'' + \Lambda\nonumber \\
{}= \frac{8\pi
\nu^2}{r^4}\Biggl[\frac{1}{2}-
\frac{q_1N''}{2}-(7q_1+4q_2+q_3)\left(\frac{N'}{r}-
\frac{2N}{r^2}\right)\nonumber\\
{}+ \frac{2(q_1+q_2+q_3)}{r^2}\Biggr]\,, \label{G3}
\end{gather}
where a
prime denotes a derivative with respect to the radial
variable $r$. Equation~(\ref{G3})
can be deduced from Eq.~(\ref{G2}) upon differentiation.

\section{Specific choice for the parameters
$q_1$, $q_2$, and $q_3$ and the reduced master equation}
\label{specific}

\subsection{Specific choice for the parameters
$q_1$, $q_2$, and $q_3$}

Clearly, Eqs.~(\ref{G1}) and (\ref{G2})
coincide when  $13q_1+4q_2+q_3 = -(7q_1+4q_2+q_3)$, i.e.,
\begin{equation}
10q_1+4q_2+q_3=0 \,. \label{q1231}
\end{equation}
This is our first choice for a relation between
the parameters $q_1$, $q_2$, and $q_3$.

We search for solutions with $N(0)=1$ and $N^{\prime}(0)=0$.
Equation~(\ref{G1}) is compatible with these conditions at $r
\to 0$, when $13q_1+4q_2+q_3 = q_1+q_2+q_3$, i.e.,
$12q_1+3q_2 = 0$, or
\begin{equation}
4q_1+q_2=0 \,. \label{q1232}
\end{equation}
This is our second choice for another relation between
the parameters $q_1$, $q_2$, and $q_3$.

These two requirements, Eqs.~(\ref{q1231}) and (\ref{q1232}),
restrict the number of the three nonminimal
coupling constants, $q_1$, $q_2$, and $q_3$, to just one
independent coupling constant, $q$, say.

We then put
\begin{equation}
q_1 \equiv -q \,, \label{q1-q}
\end{equation}
and so
\begin{equation}
q_2=4q\,, \label{q24q}
\end{equation}
\begin{equation}
q_3=-6q\,. \label{q3-6q}
\end{equation}
We assume that $q>0$.

\subsection{Reduced master equation}

With these choices
the remaining master equation for the gravitational field
(\ref{G1}) can be rewritten as,
\begin{equation}
\frac{d}{dr} \left[r (N-1)\left(1+\frac{{2Q_{m}^2} q}{r^4}\right)
\right] =  - \frac{{Q_{m}^2}}{r^2} -\Lambda r^2 \,, \label{Nn}
\end{equation}
where
\begin{equation}
{Q_{m}^2}\equiv 4\pi\nu^2 \,, \label{magneticcharge}
\end{equation}
with $Q_m$ being the magnetic charge.

\section{Exact solutions to the gravitational
field equations: generic analysis and horizons}
\label{generic}

\subsection{Four-parameter family of exact solutions with
$q_1=-q$, $q_2=4q$, and $q_3=-6q$: A
preliminary generic analysis}\label{sec3a}

The solution to Eq.~(\ref{Nn}) is of the form
\begin{equation}
N= 1 +
\left(\frac{r^4}{r^4+{ 2Q_{m}^2} q}\right)
\left(-\frac{2M}{r}
+ \frac{{Q_{m}^2}}{r^2} -\frac{\Lambda}{3}r^2 \right)\,.
\label{N00}
\end{equation}
Thus, we deal with a four-parameter family of exact solutions.
The first parameter is the nonminimal
parameter of the theory $q$,
the second parameter
is the cosmological constant $\Lambda$,
inbuilt into the theory,
the third parameter,
$Q_{m}$, is the
magnetic charge of the Wu-Yang gauge field,
and the
fourth parameter,
$M$, is the asymptotic mass of the object,
that makes its appearance as a constant of integration.
We consider
\begin{equation}
q>0 \,.\label{qpositive}
\end{equation}
\begin{equation}
\Lambda>0\,,\quad \Lambda=0\,,\quad \Lambda<0 \,,\label{landarange}
\end{equation}
\begin{equation}
{Q_{m}^2}>0 \,,\label{qmrange}
\end{equation}
\begin{equation}
M\geq0\,,
\label{mgeq0}
\end{equation}
in what follows.

The limiting case, $q=0$, gives the minimal coupled theory.
The corresponding solution, taken as the limit $q=0$ of
Eq.~(\ref{N00}), is the magnetic Reissner-Nordstr\"om
solution with a cosmological constant,
\begin{equation}
N(r) = 1 - \frac{2M}{r} + \frac{Q_m^2}{r^2} -
\frac{\Lambda}{3} r^2 \,.\label{RN0}
\end{equation}
This  magnetic solution has curvature singularities at $r=0$.
In addition, for $q<0$, Eq.~(\ref{N00}) yields
spacetime curvature singularities
at finite positive $r$. For $q>0$ there are no
curvature singularities as shown below.
That is why we choose $q>0$.

Near the
center the metric function $N(r)$ behaves as follows:
\begin{equation}
N(r) =  1 + \frac{r^2}{2 q}  - \frac{Mr^3}{{Q_{m}^2} q} +
\dots \label{N002}
\end{equation}
thus displaying that $N(0)=1$, $N^{\prime}(0)=0$ and $N^{\prime
\prime}(0)= \frac{1}{q}$. This means that the point $r=0$ is a
minimum of the regular function $N(r)$ independently on the sign and
value of the cosmological constant $\Lambda$, and independently of
the mass value $M$. Since $N(0)=1$, the curvature scalar is
regular in the center: $R(0)=\frac{6}{q}$. The quadratic scalar
$R_{mn}R^{mn}= \frac{9}{q^2}$, and other curvature invariants are
also finite in the center for $q>0$ as we
are considering. Thus the spacetime is truly regular at
the center, and this is due to
the nonminimality of the model. The
cosmological constant does not contribute to the regularity at the
center. All the objects displayed are regular objects.
We can anticipate some features of
the function $N(r)$.
$N(r)$,
independently of the parameters $q$, $\Lambda$, $Q_{m}$, and $M$,
has a central small cavity.
Where does the
boundary of this small cavity is situated? The simplest definition of
this boundary can be associated with the nearest maximum of the
function $N(r)$. The position of
this maximum, which gives the width of the
small cavity
from $r=0$ up to this maximum, with the
corresponding value
for $N$, is predetermined by the values of all four
parameters. However, it is the nonminimal parameter
$q$ that provides the creation of
this central small cavity.
For $r$ greater than this
first maximum, there exist a
zone in which the first derivative $N^{\prime}(r)$ is negative.
Thus, we deal  with a repulsion zone here.
The size of the
repulsion zone is set by the parameter $Q_{m}$,
roughly of the order of $Q_{m}$.
In addition
when the mass $M$ is small enough and the
cosmological constant is negative, $\Lambda<0$, the maximum can be
converted into an inflection point and the repulsion barrier
vanishes.
For a wide range of values of the parameters
$q$, $M> 0$, $Q_{m}^2>0$, and $\Lambda$ there exists a
Newtonian-type attraction zone of the metric function
$N(r)$, where it behaves as $1/r$.
When $\Lambda{=}0$, this zone starts at infinity
and finishes at some point.
When $\Lambda \neq 0$, this
Newtonian attraction zone appears only if the mass of the object
exceeds some critical value.
Finally, depending on the sign and value
of the cosmological constant
the solutions are asymptotically dS,
asymptotically flat, or asymptotically AdS.

The most important feature of our
investigation is the analysis of
the horizons as functions of
the
parameters.
Cauchy, event, and cosmological horizons can appear,
and in some instances they can coincide
one with another or the three altogether.
The details are given below.

\subsection{Horizons}\label{sec3c}

\subsubsection{Location of the horizons}
When $N(r_h)=0$, and the roots $r_h$ are real and
positive, we deal with horizons at these radii.
Since the equation $N(r_h)=0$ can be reduced to an algebraic
equation of order six, see Eq.~(\ref{N00}),
$
-\frac{\Lambda r^6}{3} + r^4 -2Mr^3 +
Q_{m}^2r^2 + 2Q_{m}^2
q =0
$,
one can be sure that the number of horizons is not more than six.
We intend to show that in the model under consideration the number
of horizons cannot be more than three.
Moreover, all the existing solutions for the horizons $r_h$,
namely,
$r_{1}$, $r_{2}$ and $r_{3}$, depend on
sign and value of $\Lambda$, i.e., the
cosmological constant participates in the formation of the causal
structure of the spacetime.
Generically
the three horizons that appear are the
Cauchy horizon, the event horizon, and the cosmological horizon.
To study the causal structure of the corresponding spacetimes we
have, first, to count the horizon number, second, to classify
the horizons, and third, to visualize them in figures.

Thus in order to find the location of the horizons, i.e.,
to find $r_h$ explicitly
one should solve
\begin{equation}
N(r) =  0\,,
\label{revNb0}
\end{equation}
i.e.,
the sixth-order equation
\begin{equation}
-\frac{\Lambda r^6}{3} + r^4 -2Mr^3 +
Q_{m}r^2 + 2Q_{m}^2
q =0\,.
\label{revNb}
\end{equation}

\subsubsection{Number of horizons}

\begin{figure*}[t]
\halign{\hfil#\hfil&\;\hfil#\hfil&\;\hfil#\hfil&\;\hfil#\hfil\cr
\includegraphics[height=3.35cm]{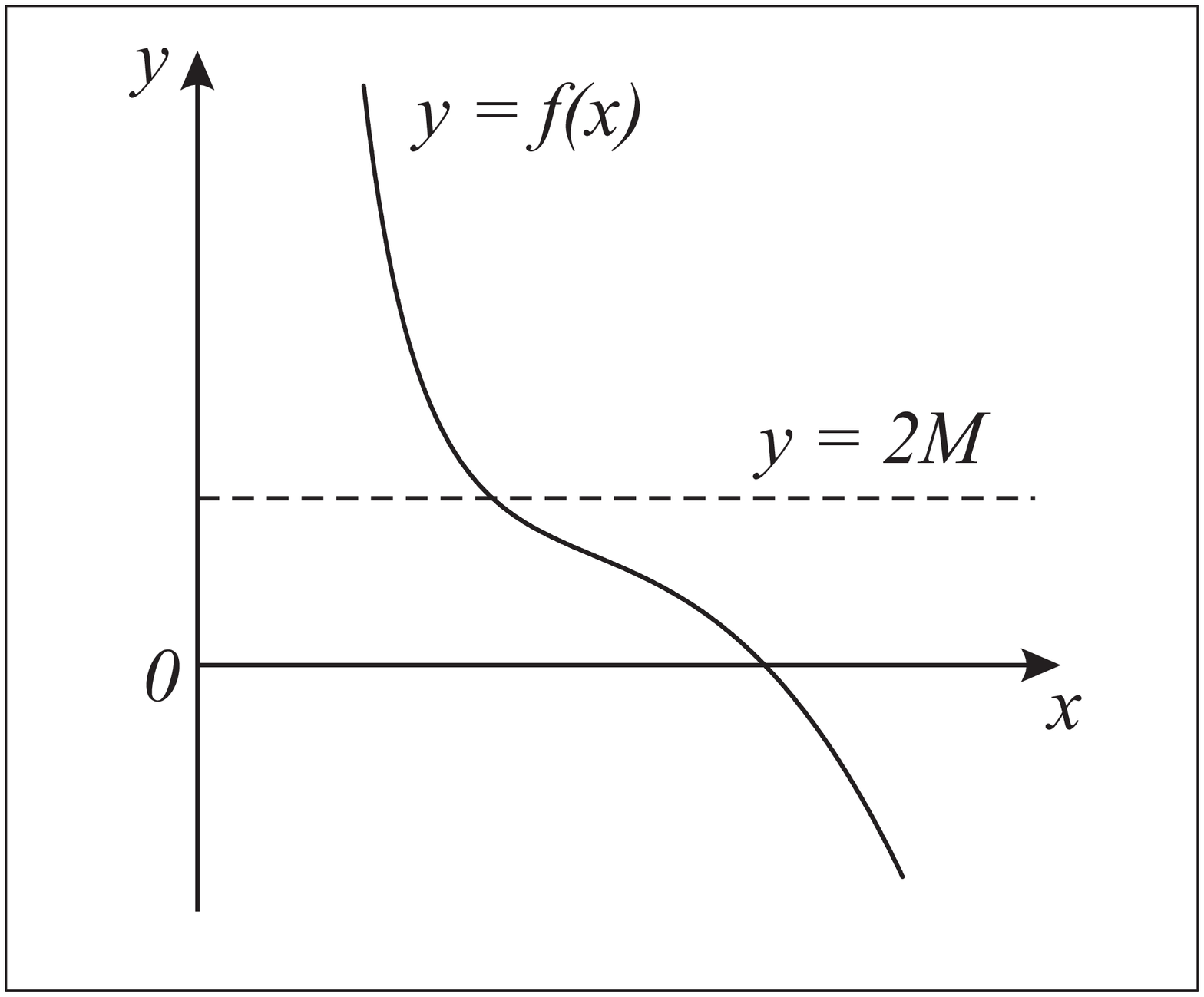}
&\includegraphics[height=3.35cm]{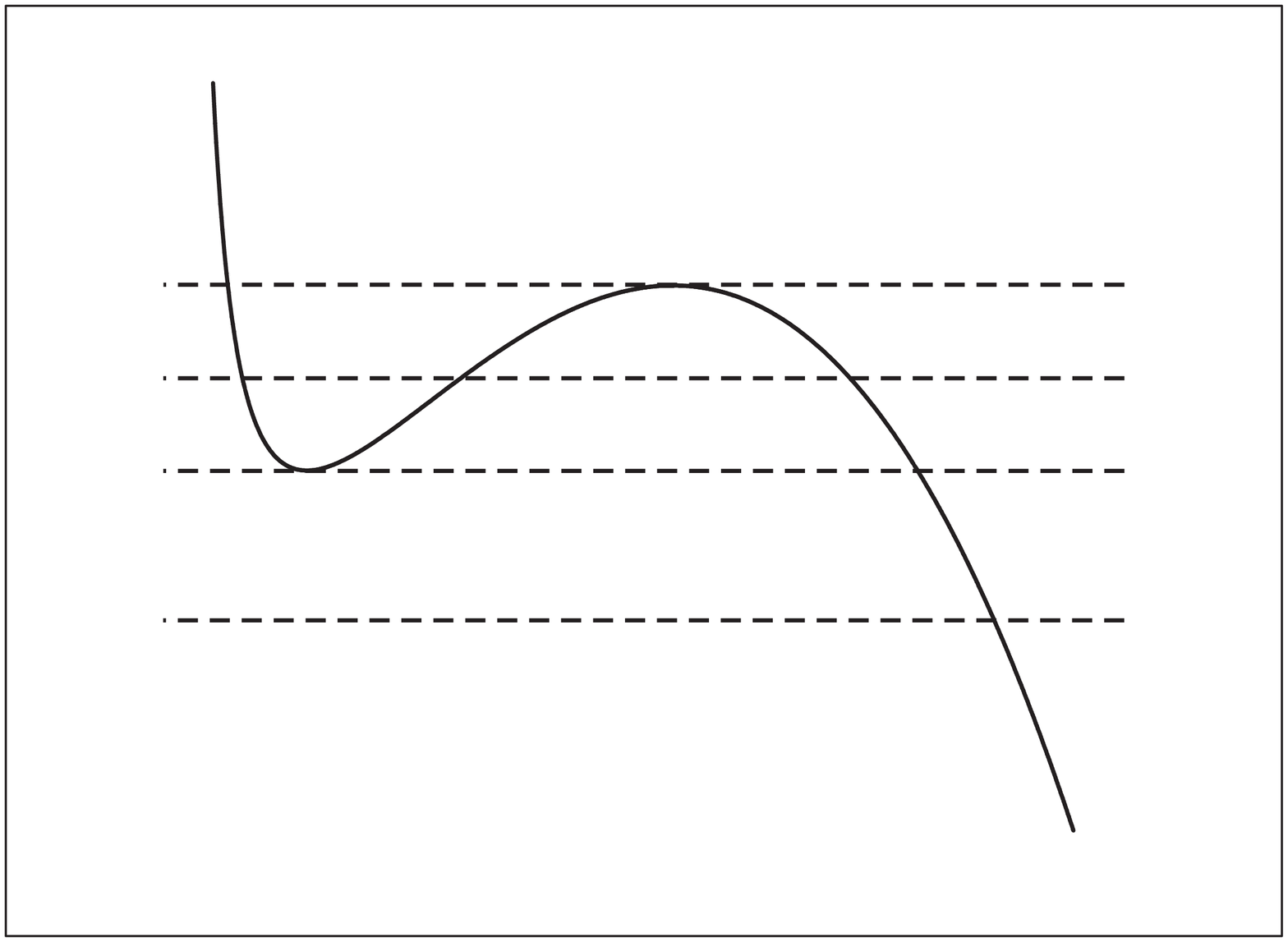}
&\includegraphics[height=3.35cm]{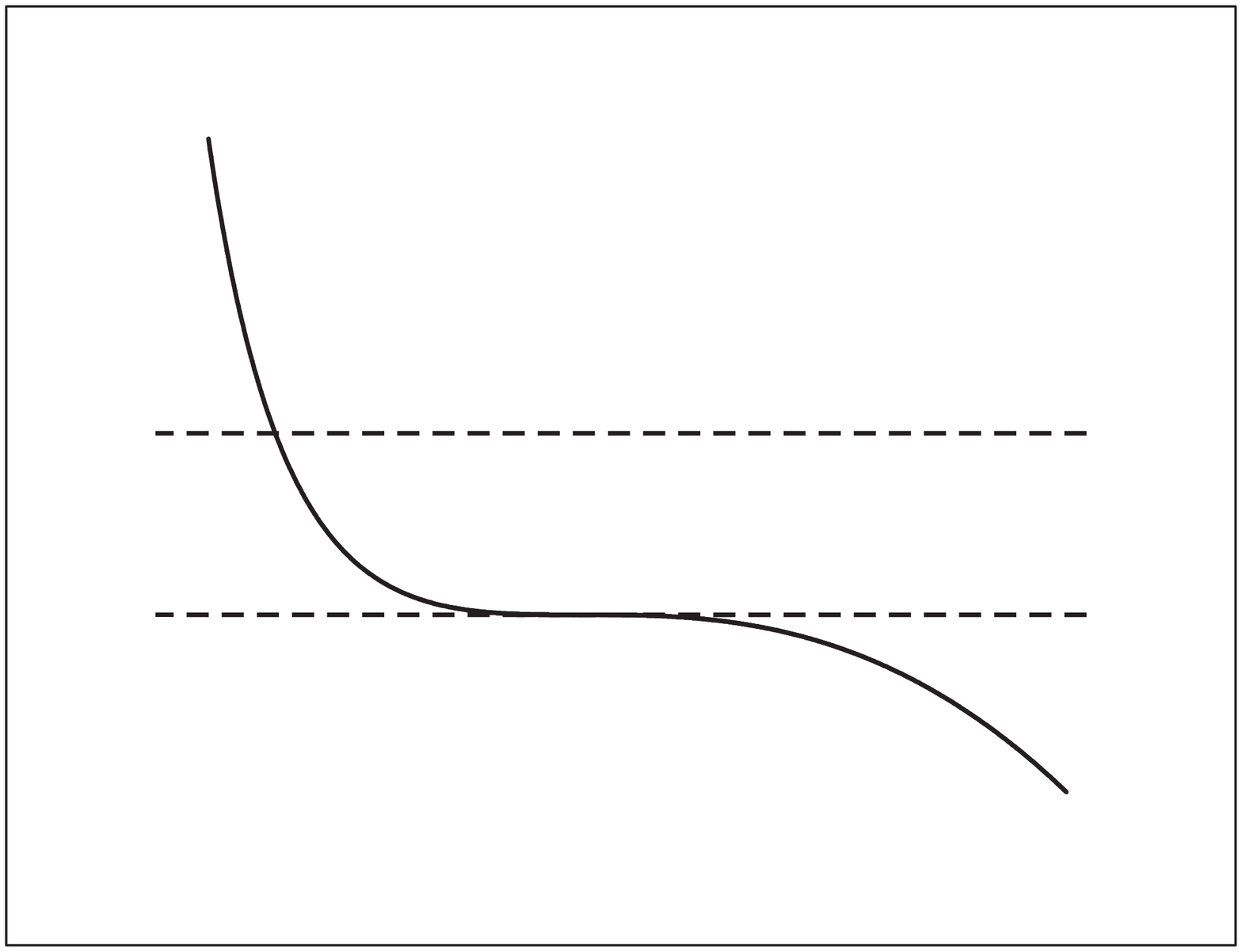}&
\includegraphics[height=3.35cm]{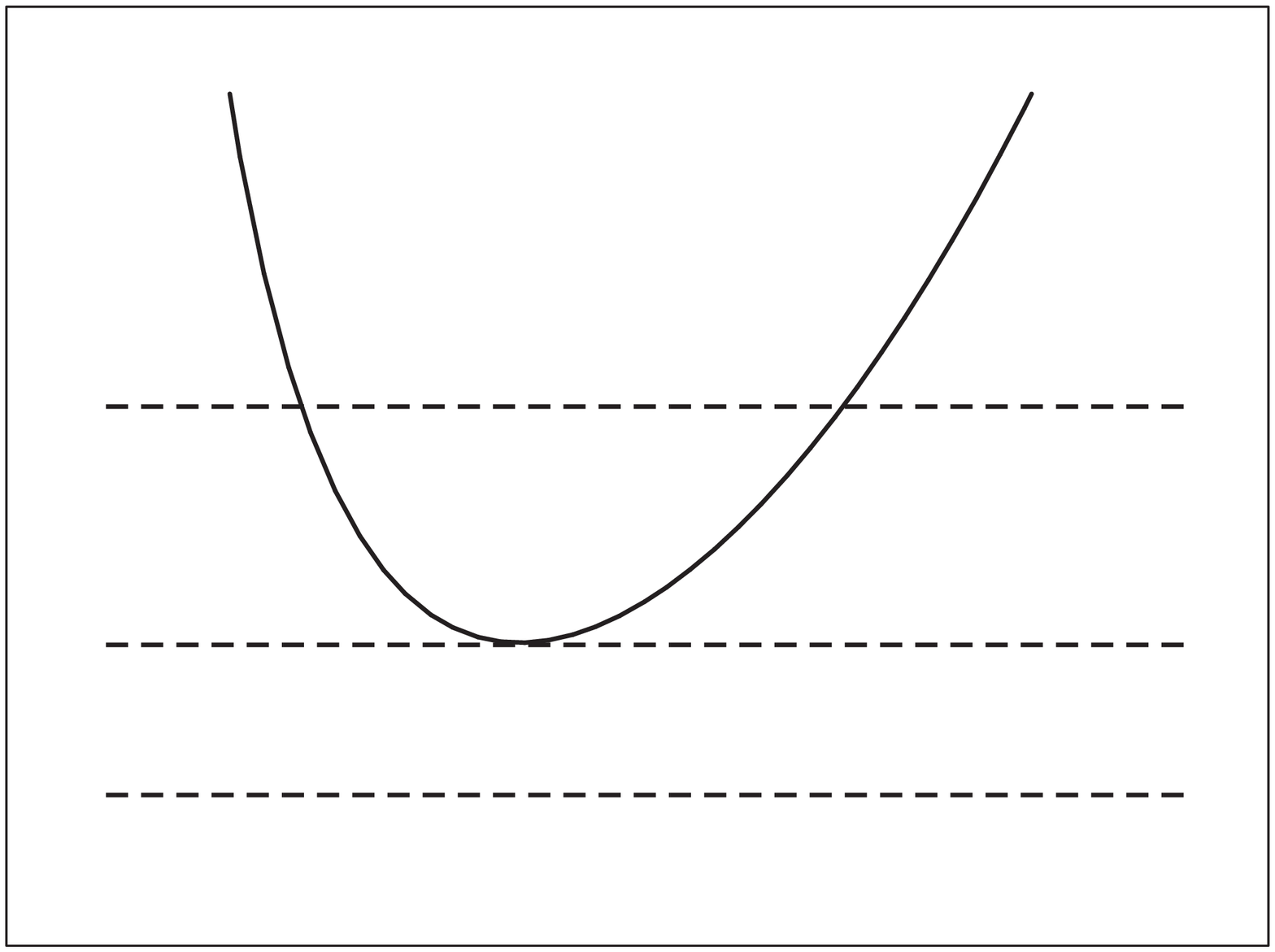}\cr
$(a)$&$(b)$&$(c)$&$(d)$\cr}
\caption{Sketches of the auxiliary function
$f(r,q,\Lambda,Q_{m}) =
-\frac{1}{3}\Lambda r^3 + r + Q_{m} r^{-1} +
{2Q_{m}^2} q r^{-3}$. In the points $x=r_h$, in which the
horizontal mass line $y=2M$ crosses the plot of the function
$y=f(r,q,\Lambda,Q_{m})$, the metric function $N(r)$ takes
zero values, i.e., the spheres $r_h$ are the horizons of the regular
nonminimal black holes.  The plot of the function
$f(r,q,\Lambda,Q_{m})$ depicted in panel $(a)$ has one inflection
point and has no extrema; it illustrates the model with one simple
horizon only. The plot in panel $(b)$ demonstrates one inflection
point and a pair of extrema (the minimum is first, the maximum is
second); depending on the mass $M$ one can obtain one or three simple
horizons, or one simple and one double horizons (the double horizon
appears when the mass-line $y=2M$ touches the plot in an
extremum). The plot in panel $(c)$ illustrates the case, when the
minimum, the inflection point and the maximum of the plot coincide;
there is only one horizon in this case, and it is triple one.
Panel $(d)$ illustrates the cases when
$f(r,q,\Lambda,Q_{m})$ has no inflection points and has only one
extremum, the minimum; in this case there are zero, one double horizon
or two simple horizons.}
\label{figc}
\end{figure*}

Another important property is the number
of the horizons.
Instead of using Eqs.~(\ref{revNb0})-(\ref{revNb})
and seeing how many horizons there are by brute
force,
one can use
a more efficient method by introducing one auxiliary
function $f(r,q,Q_{m},\Lambda)$ in the following context.
Indeed, Eq.~(\ref{revNb})
can be rewritten in the form
\begin{equation}
2M = f(r,q,\Lambda,Q_{m}) \,,
\label{revN00d2M}
\end{equation}
\begin{equation}
f(r,q,\Lambda,Q_{m})
\equiv  -\frac{\Lambda r^3}{3} +\, r\, + \frac{Q_{m}^2}{r} +
\frac{{2Q_{m}^2} q}{r^3} \,.\label{revN00d}
\end{equation}
To count the horizon number we have to determine the number of points
in which the plot of the function $y=f(r,q,\Lambda,Q_{m})$ is
crossed by the horizontal line $y=2M$, the mass line.  From a physical
point of view this procedure shows how many horizons can appear when
the mass of the object is equal to a given $M$. There are only four
types for the function $f(r,q,\Lambda,Q_{m})$. They are
depicted in Fig.~\ref{figc}.

We consider the cosmological constant $\Lambda$ to be positive, zero,
and negative. It is reasonable to start with the analysis of the case
$\Lambda>0$, i.e., asymptotically dS,
which is the most interesting and complex.  Then
we study the case
$\Lambda=0$,  i.e., asymptotically flat,
in order to recall the partially known results.
Finally, we further analyze the case
with $\Lambda<0$, i.e., asymptotically AdS.

\section{Regular nonminimal black holes with positive
cosmological constant, $\Lambda>0$}
\label{g0}

\subsection{Introduction}

Let us consider spacetimes with dS asymptotics.
The model with $\Lambda>0$ has a variety
of structures since a
cosmological horizon appears in addition to
the usual inner and outer
horizons. All spacetimes are asymptotically dS.

\subsection{Number of horizons and critical masses}

\subsubsection{Number of horizons}

We start with
the analysis of the number of horizons.
We focus on Eqs.~(\ref{revN00d2M})-(\ref{revN00d})
using the form
\begin{equation}
2M = f(r,q,|\Lambda|,Q_{m}) \,,
\label{revN00-1}
\end{equation}
\begin{equation}
f(r,q,|\Lambda|,Q_{m}) \equiv -\frac{|\Lambda| r^3}{3} + r +
\frac{Q_{m}^2}{r} +
\frac{{2Q_{m}^2}}{r^3} q \,,\label{1revN}
\end{equation}
where we have written $\Lambda=|\Lambda|$ in order
to make explicit that the cosmological constant
is positive.
Clearly, the auxiliary function $f(r,q,|\Lambda|,Q_{m})$
takes negative infinite values at $r \to \infty$, and
positive infinite values at $r \to 0$.

The extrema of $f(r,q,|\Lambda|,Q_{m})$
are given through the equation
\begin{equation}
\frac{d}{dr}f(r,q,|\Lambda|,Q_{m})=0\,,
\label{pointsofextremalambda>0}
\end{equation}
which in turn from Eq.~(\ref{1revN})
yields the  equation
\begin{equation}
|\Lambda| r^6 -r^4 + Q_{m}^2 r^2
+ 6 {Q_{m}^2} q =0 \,.
\label{revN010}
\end{equation}
In terms of the auxiliary variable $X
\equiv r^2$, Eq.~(\ref{revN010}) gives the cubic equation
$|\Lambda| X^3 -X^2 + Q_{m}^2 X + 6 {Q_{m}^2} q =0$. It has
three roots $X_{1}$, $X_{2}$, and $X_{3}$,
for which the product
$X_{1} X_{2} X_{3} = - \frac{6Q_{m}^2}{|\Lambda|}q $ is
negative, and the sum $X_{1} + X_{2} + X_{3} =
\frac{1}{|\Lambda|}$ is positive. There
are three cases:
(i) there are three real roots and two of them are
positive, (ii) there are two coinciding positive roots and one
negative, and (iii) one real root is negative and other two roots are
complex conjugated.
This means that three
different situations are
available. The configuration can have
two extrema (one maximum and one minimum), or it can have two
coinciding extrema, or still it can have
no extrema.

The points of inflection of $f(r,q,|\Lambda|,Q_{m})$
are given through the equation
\begin{equation}
\frac{d^2}{dr^2} f(r,q,|\Lambda|,Q_{m})=0\,,
\label{pointsofinfllambda>0}
\end{equation}
which in turn from Eq.~(\ref{1revN}) yields the bicubic equation
\begin{equation}
-|\Lambda| r^6 + {Q_{m}^2} r^2 + 12 {Q_{m}^2} q =0 \,.\label{1revN01}
\end{equation}
In terms of the auxiliary variable $Y
\equiv r^2$, Eq.~(\ref{1revN01}) gives the cubic equation
$-|\Lambda| Y^3 + {Q_{m}^2} Y
+ 12 {Q_{m}^2} q =0$. It has
three roots $Y_{1}$, $Y_{2}$, and $Y_{3}$,
for which the product $Y_{1} Y_{2}
Y_{3} = \frac{12{Q_{m}^2}}{|\Lambda|}q$ is positive,
and the sum is
equal to zero, $Y_{1} + Y_{2} +  Y_{3} =0 $. This means that
only two cases are available: (i) two real roots are negative
and one real root is positive, and (ii)
the real part of a complex
conjugated pair of roots is negative and the only real root is
positive.
Since negative and complex roots are not physical, we conclude
that there exists only one inflection point of
the graph of the function $f(r,q,|\Lambda|,Q_{m})$.

The sketches of
the corresponding three plots of the function
$f(r,q,|\Lambda|,Q_{m})$ are presented in Fig.~\ref{figc},
panels (a), (b) and
(c). The mass line $y\equiv 2M$  crosses the corresponding
plots at least in one point, meaning that
there is at least one
horizon. More precisely, depending on the values of parameters
$q$, $M$, $Q_{m}$,
and $|\Lambda|$,
we deal with three independent simple horizons, or with one
simple and one double horizon, or with one triple horizon,
or even with one simple
horizon.

\subsubsection{The critical masses}

The
analysis of
horizons and extrema shows that there are two critical masses,
$M_{\rm c1}$ and $M_{\rm c2}$, say, for which the
corresponding horizons coincides with one of the extrema.
For these masses the horizon is in a fact a double
horizon, it is degenerate.

One coincidence
takes place when the plot of the function $N(r)$ touches the line
$N=0$ in its minimum, and another coincidence
happens  when $N(r)$ touches  the line
$N=0$ in its maximum. The critical masses can be calculated as
\begin{equation}
M_{\rm c1} = \frac12 f(r_{\rm c1},q,|\Lambda|,Q_{m}) \,, \quad
\label{D170}
\end{equation}
\begin{equation}
M_{\rm c2} = \frac12 f(r_{\rm c2},q,|\Lambda|,Q_{m})\,,\label{D17}
\end{equation}
where the radii $r_{\rm c1}$ and $r_{\rm c2}$ are obtained as the
corresponding solutions of the pair of equations $N(r)=0$ and
$\frac{dN(r)}{dr}=0$.

Thus, two critical parameters $M_{\rm c1}$ and
$M_{\rm c2}$ appear in the course of analysis.
These critical masses can also
coincide. In this case we deal with a triple
horizon, when the
two extrema, the inflection point and all three zeros
of the function $N(r)$ coincide.

\subsection{The function $N$ and the  eight distinct cases}

\subsubsection{The function $N$ in terms of auxiliary parameters
and effective parameters}

The family of solutions under discussion is described by four
positive parameters with different
intrinsic units:  $q$, $M$, $Q_{m}$,
and $\Lambda$. In order to clarify
the fine details of these
solutions we use below auxiliary and
effective parameters and
auxiliary and effective variables.
First, we introduce
four auxiliary parameters with units of length
\begin{gather}
r_{\Lambda} \equiv \sqrt{\frac{3}{|\Lambda|}} \,, \quad r_{g}
\equiv 2M \,, \nonumber \\
r_{q} \equiv ({2Q_{m}^2} q)^{\frac{1}{4}} \,,
\quad r_{Q} \equiv Q_m \,.\label{HOR3}
\end{gather}
Second, we construct three effective positive
parameters without units, namely,
\begin{equation}
\alpha \equiv \frac{r_{g}}{r_{\Lambda}} \,, \quad \beta \equiv
\left(\frac{r_{Q}}{r_{\Lambda}}\right)^2  \,, \quad  \gamma
\equiv \left(\frac{r_{q}}{r_{\Lambda}}\right)^4 \,.\label{HOR4}
\end{equation}
Third, since $\Lambda \neq 0$, in fact here
$\Lambda> 0$, we introduce the
variable $x$ without units through
\begin{equation}
x \equiv \frac{r}{r_{\Lambda}}\,.
\label{}
\end{equation}
We thus use the
cosmological scale for the description of
the internal structure of
the regular nonminimal black holes.

In  terms of these variables
the metric $N(r)$, given in Eq.~(\ref{N00}),
is now $N(x)$ and takes the form
\begin{equation}
N(x)= \frac{\gamma+ \beta x^2 - \alpha x^3 + x^4 -x^6}{x^4+\gamma}
\,.\label{D11}
\end{equation}
Its first derivative is
\begin{equation}
N^{\prime}(x)=-\frac{x\left[2x^8 - \alpha x^5 +
2x^4(\beta+3\gamma)+ 3\alpha \gamma x - 2\beta
\gamma \right]}{(x^4+\gamma)^2} \,,
\label{D13}
\end{equation}
where ${}^{\prime}\equiv \frac{d\,}{dx}$.
Clearly, only the three parameters without
units, $\alpha$, $\beta$, and $\gamma$, are the
ones that matter
in the search for horizons, $N(x)=0$, and  in the
search for extrema, $N^{\prime}(x)=0$.

\begin{figure*}[t]
\halign{\qquad\hfil#\hfil&\;\hfil#\hfil\cr
\includegraphics[height=6cm]{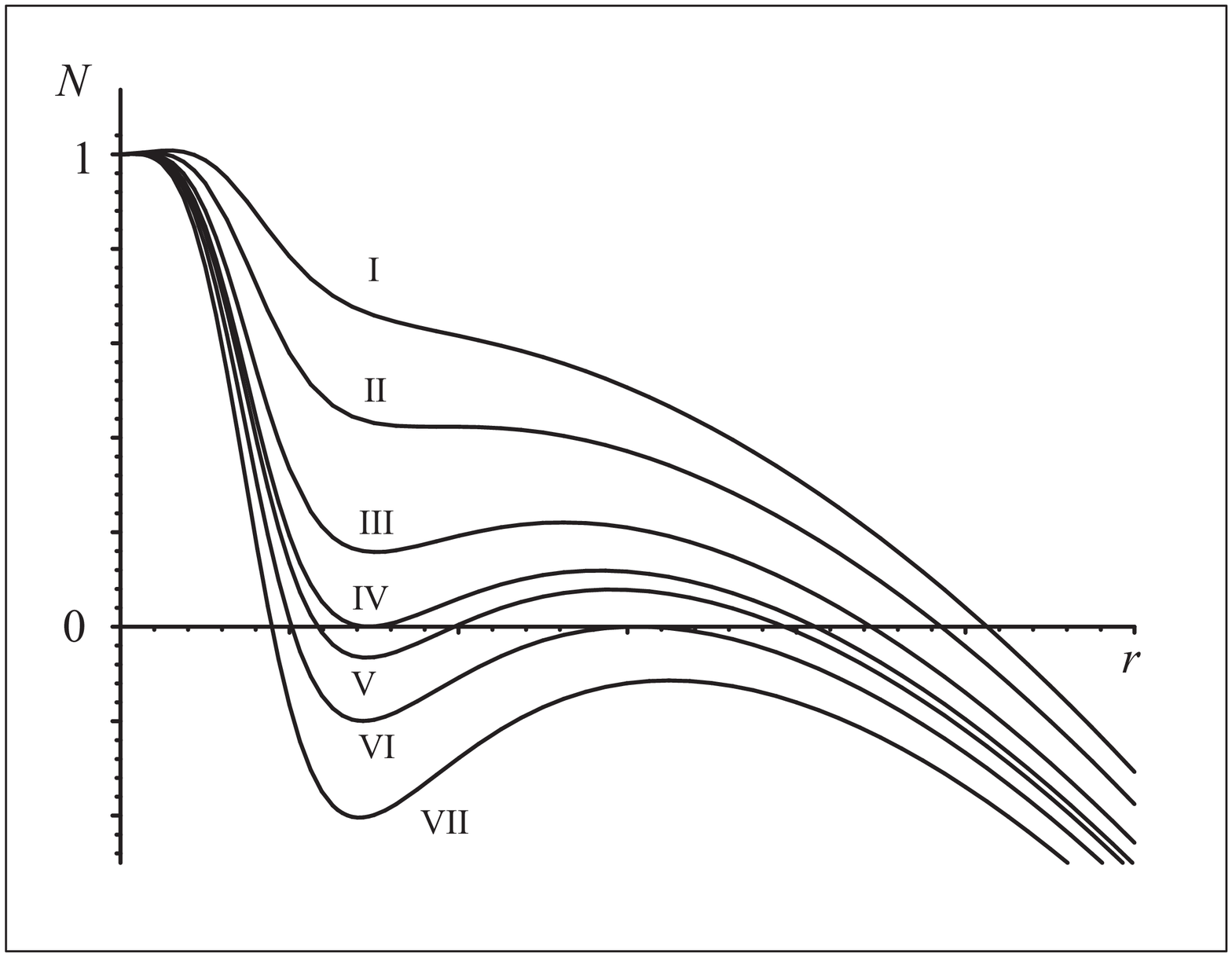}&
\includegraphics[height=6cm]{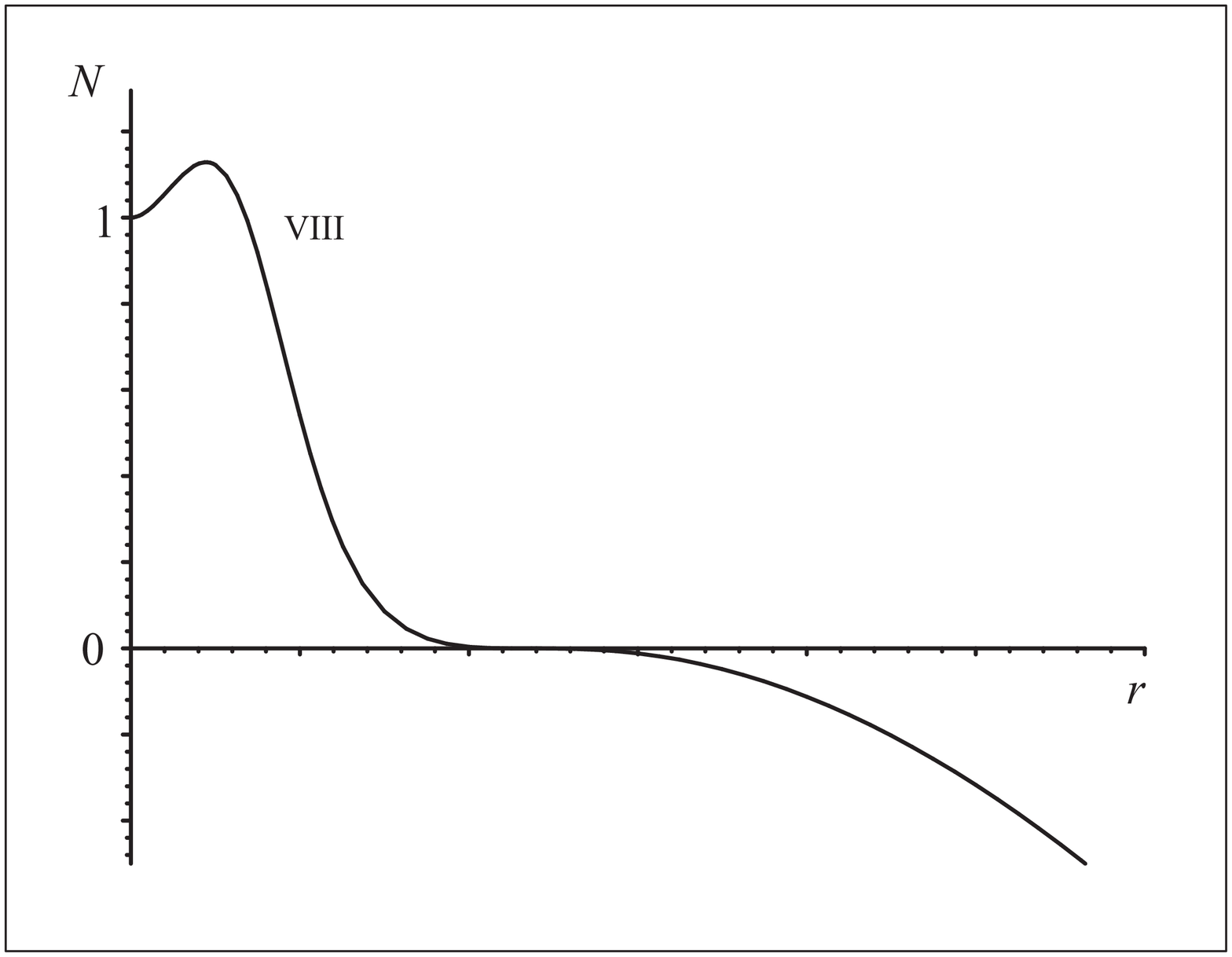}\cr
$a$&$b$\cr} \caption{$\Lambda>0$.
The plots of the metric function $N(r)$ for the nonminimal regular
models with $\Lambda>0$.  Note that $N(0)=1$ for all curves
so that the spacetimes are regular.
All the curves have small cavities near the
center, repulsive barriers, and a dS asymptotic behavior.  In panel
$(a)$, seven cases are displayed.  Curve I is the case $M=0$.  Curve II
is representative of the case $0<M\leq M_{\rm N}$ where there is no
zone of Newtonian-type attraction.  Curve III, for $M_{\rm N}<M<M_{\rm
c1}$, describes regular nonminimal stars inside a cosmological
horizon, and possessing a Newtonian attraction zone.  Curve IV
represents the case $M=M_{\rm c1}$, where $M_{\rm c1}$ is a critical
mass.  There is an extremal nonminimal regular black hole, with the
Cauchy and event horizons coinciding, it is a double horizon.  Curve V
is for $M_{\rm c1}<M<M_{\rm c2}$. Here the nonminimal regular black
hole has a Cauchy horizon, an event horizon and a distant cosmological
horizon.  Curve VI is the case $M=M_{\rm c2}$, where $M_{\rm c1}$ is
another critical mass.  The event and cosmological horizons coincide,
so it is a double horizon.  Curve VII for $M_{\rm c2}<M$ represents
ultramassive black holes with a regular center and a Cauchy horizon
that has turned into a cosmological horizon.  In panel $(b)$, curve
VIII has a triple horizon of the cosmological type.  For this curve
four specific points coincide: the inflection point and three zeros of
the function $N(r)$.
}
\label{fig5}
\end{figure*}

In this analysis the masses
$M_{\rm c1}$ and $M_{\rm c2}$
given in Eqs.~(\ref{D170})-(\ref{D17})
are important. They
give the cases in which there is a double horizon.
This double horizon can be
an extreme black hole horizon
or an extreme cosmological horizon,
each with a corresponding
radius
$x_{\rm c1}$ and $x_{\rm c2}$.
The critical masses can
coincide giving a triple
horizon.
The plots of $N(r)$
are presented in Fig.~\ref{fig5}.
Below we describe the
features of the solutions for
the different characteristic masses.

\subsubsection{The eight distinct cases}

\noindent {(i) The case $M=0$}

\noindent
When the mass $M$ vanishes, we deal with a regular
nonminimal object that is not a black hole. At
$x=0$ there is a minimum with $N(0)=1$; then the function $N(x)$
reaches a maximum at $x=x_{{\rm max}}$, where
\begin{equation}
x_{{\rm max}}= \left[\sqrt{\frac14 (\beta+3\gamma)^2 + \beta
\gamma} -\frac12 (\beta+3\gamma) \right]^{\frac14} \,.
\label{D19}
\end{equation}
The zone $0<x<x_{{\rm max}}$ is a central
attraction zone. The zone
$x_{{\rm max}}<x<\infty$
is a repulsion zone. There is one
inflection point of curve $N(x)$ in this zone.

This
spacetime has only one horizon, namely, the cosmological
horizon at
$r_{\rm ch}=\sqrt{\frac{3}{|\Lambda|}} x_{\rm ch}$,
where $x_{\rm ch}$ can
be found as follows. Define $\gamma_{*}$ and $\Gamma$ as
\begin{equation}
\gamma_{*}=\frac{2}{27}\left[(1+3\beta)^{\frac32}-1\right]
-\frac13 \beta\,,
\label{N13u0}
\end{equation}
\begin{equation}
\Gamma = 1 +
\frac{27(\gamma-\gamma_*)}{2(1+3\beta)^{\frac32}} \,.\label{N13u}
\end{equation}

\noindent When $\gamma>\gamma_{*}$, one obtains that
\begin{align}
x_{\rm ch}&= \sqrt{\frac13 + {\bar \chi}} \,, \nonumber\\
{\bar \chi} &= \frac23
\sqrt{1+3\beta} \ \cosh{\left[\frac13
\log{\left(\Gamma+\sqrt{\Gamma^2-1}\right)}\right]} \,.\label{D21}
\end{align}

\noindent When $\gamma=\gamma_{*}$, one obtains that
\begin{equation}
x_{\rm ch}= \sqrt{\frac13 + \chi_{*}} \,, \quad \chi_{*} = \frac23
\sqrt{1+3\beta} \,.\label{D31}
\end{equation}

\noindent When $\gamma<\gamma_{*}$, one obtains that
\begin{align}
x_{\rm ch}&= \sqrt{\frac13 + \tilde \chi} \,, \nonumber\\
{\tilde \chi}&= \frac23
\sqrt{1+3\beta} \ \cos{\left(\frac13 \arccos{\Gamma}\right)} \,.
\label{D27}
\end{align}
This case is illustrated by curve I in Fig.~\ref{fig5}.

\vspace{3mm} \noindent {(ii) The case $0<M\leq M_{\rm
N}$}

\noindent
Inside the range $0<M<M_{\rm c1}$
there exists a specific mass value,
\begin{equation}
M_{{\rm N}}=\frac12
f(r_{{\rm N}},q,|\Lambda|,Q_{m})\,,
\end{equation}
which distinguishes the models with no
Newtonian-type attraction zone
from the ones with a
Newtonian-type attraction zone, i.e.,
with a
$1/r$ dominating term in the gravitational potential.
For masses in the range $0<M\leq M_{\rm
N}$ there is no Newtonian-type attraction zone.
The corresponding radius
$x_{{\rm N}}$ appears as a solution of the pair of equations,
$N^{\prime}(x_{{\rm N}})=0$ and $N^{\prime \prime}(x_{{\rm
N}})=0$; i.e., this point of the graph of the function $N(x)$
has a coincidence of extrema points and an inflection point.
This limiting
case, where the mass is $M_{{\rm N}}$
is illustrated by curve II in Fig.~\ref{fig5} and
is representative of the range  $0<M\leq M_{N}$.
The nonminimal regular objects within this
mass range
are not black holes, since the single simple horizon
is a cosmological horizon.

\vspace{3mm} \noindent {(iii) The case $M_{\rm
N}<M<M_{\rm c1}$}

\noindent
Clearly, within this range the following
features can be visualized, see
curve III in Fig.~\ref{fig5}. There is, first,
a central small cavity (between the center $r=0$ and the nearest
maximum), second, a
repulsion barrier (between the nearest
maximum and the following minimum), third, the zone of
Newtonian-type attraction, i.e., similar
to the $1/r$ behavior of
the Newtonian gravitational potential
(between the minimum and the final
maximum), and fourth,
the zone of cosmological repulsion
(between the final
maximum and infinity).
The nonminimal regular objects within this
mass range
are not black holes, since the single simple horizon
is a cosmological horizon.

\vspace{3mm} \noindent {(iv) The case $M_{\rm c1}=M$}

\noindent
When the mass $M$ has this value, the nonminimal
coupled regular black
hole is extremal with the inner and the even horizon
coinciding. The minimum
coincides
with a double horizon.
There is still an outer cosmological horizon.
This is shown in curve
IV of Fig.~\ref{fig5}.

\vspace{3mm} \noindent {(v) The case $M_{\rm c1} <
M < M_{\rm c2}$}

\noindent
When the mass $M$ is in this range, we deal with
regular nonminimal black holes inside a cosmological horizon.
Specific features of these solutions are the following, see the curve
V in Fig.~\ref{fig5}.
Each solution of this type displays
a small cavity
near the regular center, a repulsion barrier, a Cauchy
horizon, a black hole region, an event horizon, a zone of
Newtonian-type attraction of $1/r$
plus corrections, an intermediate maximum, and a zone of
cosmological repulsion.

\vspace{3mm} \noindent {(vi) The case $M_{\rm c2}=M$}

\noindent
When the mass $M$ has this value, the nonminimal
coupled regular black
hole is extremal with the even horizon
coinciding with the cosmological horizon.
The maximum
coincides with a double horizon.
There is still an inner Cauchy horizon.
This is shown in curve
VI of Fig.~\ref{fig5}.
In this case the black hole is a cosmological extremal
supermassive regular black hole as the whole visible universe is
swallowed by this supermassive object.

\vspace{3mm} \noindent {(vii) The case $M_{\rm c2}<M$}

\noindent
When the mass exceeds the second critical mass, the
Cauchy horizon of the regular nonminimal black hole becomes the
cosmological horizon. We deal now with a supermassive black
hole with central small cavity, a
unified repulsion barrier, and without
a standard zone of Newtonian-type attraction, see curve VII on
Fig.~\ref{fig5}.
Since for $ M>M_{\rm c2}$ the Cauchy
horizon is also a cosmological horizon, the black hole is
ultramassive. In such universe there is only one horizon, which is
a cosmological one, together with a repulsion zone. In this context this
ultramassive black hole is of a new type, for which all three
horizons coincide,
namely, the Cauchy, event and the cosmological horizons.

\vspace{3mm} \noindent {(viii) The case $M_{\rm c1}=
M_{\rm c2}$}

\noindent When the critical masses $M_{\rm c1}$ and $M_{\rm c2}$
coincide, two separatrices convert into one curve, depicted in panel
(b) of Fig.~\ref{fig5}, curve VIII.
This curve illustrates the situation, when four
specific points of the function $N(x)$ coincide,
namely, the maximum, the minimum,
the inflection point and the horizon. We deal now with a triple
horizon. When $M< M_{\rm c1}=M_{\rm c2}$, the solutions behave
like curve III (but without extrema). When $M>M_{\rm
c1}=M_{\rm c2}$ the curves look like the curve VII (but without
extrema).

\section{Regular nonminimal black holes with zero cosmological
constant, $\Lambda=0$}
\label{eq0}

\subsection{Introduction}

This case has been studied in \cite{BaZaPLB}
and we briefly review the results.
All spacetimes are asymptotically flat.

\subsection{Number of horizons and the critical masses}

In this case we do not need to
introduce variables without units as we
did in the previous case.  A typical sketch of  the auxiliary
function $f(r,q,0,Q_{m})$ looks like the curve depicted in
panel (d) of Fig.~\ref{figc}.  There is a critical mass $M_{\rm c}$,
corresponding to the minimum of that curve.  Depending on the
value of $M$ of the object, the line $y=2M$ can cross the curve
$y=f(r,q,0,Q_{m})$ zero times, one time or two times. The
corresponding spacetimes have no horizons, when the mass $M$ is less
than the critical mass, $M<M_{\rm c}$, it has one horizon, a double
one, when $M=M_{\rm c}$, and it has two horizons, when $M>M_{\rm c}$.

In more detail,
for $\Lambda=0$, Eqs.~(\ref{revN00d2M})-(\ref{revN00d})
turn into
\begin{equation}
2M = f(r,q,0,Q_{m}) \,,
\label{revN00d2M0}
\end{equation}
\begin{equation}
f(r,q,0,Q_{m})= r +
\frac{{Q_{m}^2}}{r} + \frac{{2Q_{m}^2} q}{r^3}
\,. \label{revN21}
\end{equation}
When the mass line $y=2M$ touches the curve
$y=f(r,q,0,Q_{m})$ at
the point of minimum, the mass is a critical mass,
$M_{{\rm c}}$, given by
\begin{equation}
M_{{\rm c}} = \frac12 f(r_{\rm c},q,0,Q_{m})\,,
\label{M10}
\end{equation}
where $r_{\rm c}$ is the critical radius.
This radius can be found by imposing
$N(r_{{\rm c}})=0$ and
$N^{\prime}(r_{{\rm c}})=0$, yielding
\begin{equation}
r_{\rm c}=\sqrt{\frac{{Q_{m}^2}}{2} +
\sqrt{\frac{{Q_{m}^4}}{8}+6{Q_{m}^2}
q}} \,.\label{D33}
\end{equation}
Equations~(\ref{M10})-(\ref{D33})
then give
\begin{equation}
M_{\rm c} = \sqrt{{2Q_{m}^2}}\,
\frac{\left(1+  \frac{8q}{{Q_{m}^2}}
+ \sqrt{1+ \frac{24q}{{Q_{m}^2}}}
\right)}{\left(1+ \sqrt{1+\frac{24q}{{Q_{m}^2}}}
\right)^{\frac{3}{2}}}\,.\label{M1}
\end{equation}
The parameter $r_{\rm c}$ is convenient for
the analyses. Using it
we obtain, in particular,
$
6{Q_{m}^2} q = r_{\rm c}^2
\left(r_{\rm c}^2- {Q_{m}^2} \right)$ and
$
M_{\rm c} = \frac{1}{3r_{\rm c}} (2r_{\rm c}^2+{Q_{m}^2})
$.

\subsection{The function $N$ and the four distinct cases}

\begin{figure}[t]
\includegraphics[height=6cm]{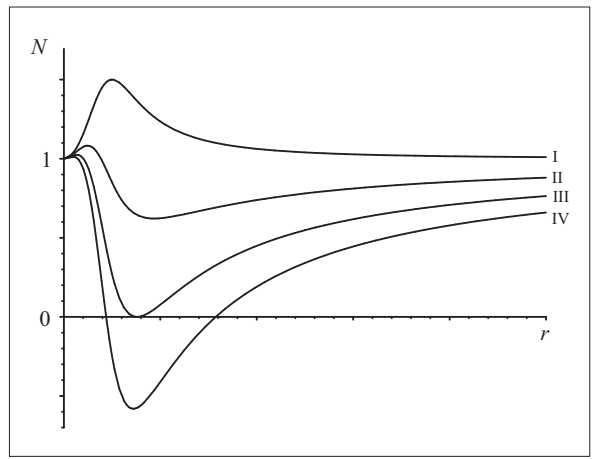}
\caption{$\Lambda=0$.
The plots of the metric function $N(r)$ for the nonminimal regular
models with $\Lambda=0$.  Note that $N(0)=1$ for all curves so that
the spacetimes are regular.  All the curves contain a small cavity
near the center, a repulsion barrier, and a flat asymptotic behavior.
Curve I displays the behavior of $N(r)$, when $M=0$.  Curve II is for
the case $0<M<M_{\rm c}$. There are no horizons in the spacetime, the
object is a regular star.  Curve III, for case $M=M_{\rm c}$, is the
critical case. There is a double horizon at the minimum of the metric
function $N(r)$, i.e., the Cauchy and event horizons coincide.  Curve
IV describes a black hole with mass exceeding the critical mass
$M>M_{\rm c}$, it contains a Cauchy horizon and an event horizon.
Curves II, III, and IV, which correspond to massive objects, have
zones of Newtonian-type attraction.}
\label{fig3}
\end{figure}

Now we give the main features of the four distinct
types that appear for $\Lambda=0$, see
Fig.~\ref{fig3}.

\vspace{3mm} \noindent
(i) The case $M=0$

\noindent
When $M=0$ we deal with a
pure nonminimal object.
The metric function $N(r)$ given in Eq.~(\ref{N00})
is
\begin{equation}
N(r)= 1+
\frac{{Q_{m}^2} r^2}{r^4+{2Q_{m}^2} q} \,.\label{D1}
\end{equation}
and its derivative has also a simple form,
$ N^{\prime} =
\frac{{2Q_{m}^2} r}{(r^4+{2Q_{m}^2} q)^2}({2Q_{m}^2} q
- r^4)$.
The plot of the function $N(r)$, see curve I in
Fig.~\ref{fig3}, is located above the horizontal line $N=1$ and
has a maximum at $r=({2Q_{m}^2} q)^{\frac{1}{4}}$, and a minimum at
the center. Thus, a Newtonian-type attraction zone is absent now,
and there is a repulsion barrier that
prolongs itself into infinity.
There is no equilibrium point and
there are no horizons in the spacetime.

\vspace{3mm} \noindent
(ii) The case $0<M<M_{\rm c}$

\noindent
When $M<M_{\rm c}$ there are no horizons in the spacetime,
and we deal with regular stars. When the mass $M$ is
in this range
one can find the point of
minimum of $N(r)$, $r_{\rm min}$, say, which is
an equilibrium point, in which the zones of attraction and
repulsion make contact, see curve II in Fig.~\ref{fig3}. Since at
this point $N'(r_{\rm min})=0$, the gravitational force does
not act on any massive particle, i.e., a
particle can be at rest at this
point. Any small perturbation puts this particle to oscillate
near this equilibrium point.

\vspace{3mm} \noindent
(iii) The case $M=M_{\rm c}$

\noindent
In the critical case $M=M_{\rm c}$
the horizon is double and has radius
$r=r_{\rm c}$ given in Eq.~(\ref{D33}).
The metric function $N(r)$ given in Eq.~(\ref{N00})
has for this case the simple form
\begin{equation}
N(r) = \frac{(r-r_{\rm c})^2}{r^4 + {2Q_{m}^2} q}\left[r^2 +
\frac{{2Q_{m}^2}
q}{r_{\rm c}^2}\left(\frac{2r}{r_{\rm c}} + 1 \right)
\right] \,.\label{N11}
\end{equation}
The multiplier $(r-r_{\rm c})^2$ shows explicitly that the horizon at
$r=r_{\rm c}$ is a
double horizon. In other words, we deal with an extremal
regular nonminimal black hole. As $r\to\infty$
the spacetime is asymptotically flat.
The plot of the function (\ref{N11}) is presented in
Fig.~\ref{fig3} curve III. The curve contains
a zone of Newtonian-type attraction, which starts at infinity
and finishes at the horizon. The interior of this extremal
black hole is regular. It contains a zone near the center
which is
a nonminimal small cavity,
and it also contains a nonminimal repulsion barrier which
is locate between the sphere on which $N(r)$ has a maximum and
the double horizon.

\vspace{3mm} \noindent
(iv) The case $M_{\rm c}<M$

\noindent
When $M_{\rm c}<M$, the interior of the regular black hole
contains a
nonminimal small cavity and a repulsion barrier that are
separated from the exterior by a nonminimal well,
see curve IV in Fig.~\ref{fig3}. The
edge of the well nearest to the center is the Cauchy
horizon, the distant edge is the event horizon, it
separates the interior from the zone of Newtonian-type attraction
with a $\frac{1}{r}$ potential term.

\section{Regular nonminimal black holes with negative
cosmological constant, $\Lambda < 0$}
\label{l0}

\subsection{Introduction}

The case $\Lambda < 0$ has some features
similar to the $\Lambda =0$ case, with in addition
some more structure.
All spacetimes are asymptotically AdS.

\subsection{Number of horizons and the critical masses}

Let us consider spacetimes with AdS asymptotics,
$\Lambda <0$.
The case $\Lambda<0$, like the $ \Lambda =0$ case,
is simpler than the case
$\Lambda>0$ and there is no need to introduce
variables and parameters with no units.

A typical sketch of the
auxiliary function $f(r,q,-|\Lambda|,Q_{m})$
looks like the curve depicted in
panel (d) of Fig.~\ref{figc}.
In this
case there is also a
critical mass $M_{\rm c}$ related to the minimum.
Depending on the value of $M$ of the object, the line $y=2M$ can cross
the curve $y=f(r,q,-|\Lambda|,Q_{m})$
zero times, one time or two times. The
corresponding spacetimes have no horizons, when the mass $M$ is less
than the critical mass, $M<M_{\rm c}$, it has one horizon, a double
one, when $M=M_{\rm c}$, and it has two horizons, when $M>M_{\rm c}$.
Since the
spacetime is
asymptotically AdS for sufficient large
radii there is a cosmological attraction.

In more detail,
for $\Lambda <0$, Eqs.~(\ref{revN00d2M})-(\ref{revN00d})
turn into
\begin{equation}
2M = f(r,q,-|\Lambda|,Q_{m})\,,
\label{revN00****2M}
\end{equation}
\begin{equation}
f(r,q,-|\Lambda|,Q_{m})
\equiv  \frac{|\Lambda| r^3}{3} + \,r \,+
\frac{{Q_{m}^2}}{r} +
\frac{{2Q_{m}^2}}{r^3} q \,,\label{revN00****}
\end{equation}
where we have set $\Lambda=-|\Lambda|$.
Since the second derivative of the function
$f(r,q,-|\Lambda|,Q_{m})$ is positive everywhere, and
$f(0,q,-|\Lambda|,Q_{m})
= f(\infty,q,-|\Lambda|,Q_{m}) = + \infty$, a typical
sketch of this function is
like panel (d) in Fig.~\ref{figc}. This function
has features similar to the function $f(r,q,0,Q_{m})$
analyzed previously.

We can find the
critical value of the mass $M_{\rm c}$. It is
\begin{equation}
M_{\rm c} \equiv  \frac12\left[\frac{|\Lambda| r_{\rm c}^3}{3}
+ r_{\rm c} + \frac{{Q_{m}^2}}{r_{\rm c}} +
\frac{{2Q_{m}^2}}{r_{\rm c}^3} q
\right]\,,\label{1reN00}
\end{equation}
where $r_{\rm c}$ is the real positive solution of the following
bicubic equation
\begin{equation}
|\Lambda| r_{\rm c}^6 + r_{\rm c}^4 - {Q_{m}^2} r_{\rm c}^2 -
6 {Q_{m}^2} q
=0 \,.\label{1revN010}
\end{equation}
In this case there is only one real positive solution.
Indeed, if we introduce the quantity $X$ as $X \equiv
r^2$, we see that the product of the roots of
Eq.~(\ref{1revN010}),  $X_{1}X_{2}X_{3}=\frac{6{Q_{m}^2}
q}{|\Lambda|}$ is positive, and the sum $X_{1}+X_{2}+X_{3}=
- \frac{1}{|\Lambda|}$ is negative. This can
happen in three instances:
first, when there are two complex roots and one real positive root,
say, $X_{1}>0$; second, when there are three real roots, two of
them being negative, and one positive, $X_{1}>0$; third, when
$X_{2}=X_{3}<0$, and $X_{1}>0$. In any case, there is only
one real positive solution $r_{\rm c}=\sqrt{X_{1}}$ of
Eq.~(\ref{1revN010}). Technically, in each of the three mentioned
above situations, the unique real positive root $X_{1}$ can be
extracted from the three corresponding solutions of the standard
Cardano formula, namely,
\begin{gather}
X= Y-\frac{1}{3|\Lambda|} \,, \quad Y_{1}= \mu + \nu\,,\nonumber\\
Y_{2,3} = -\frac12(\mu + \nu) \pm i \frac{\sqrt3}{2}(\mu -
\nu)\,,\label{cardano1}
\end{gather}
where
\begin{gather}
\mu \equiv \left[-\frac{{\cal Q}}{2} + \sqrt{\frac{{\cal P}^3}{27}
+ \frac{{\cal Q}^2}{4}} \right]^{\frac13}\,, \nonumber\\ \nu \equiv
\left[-\frac{{\cal Q}}{2} - \sqrt{\frac{{\cal P}^3}{27} +
\frac{{\cal Q}^2}{4}} \right]^{\frac13}\,,\label{cardano2}
\end{gather}
\begin{equation}
{\cal Q} \equiv \frac{2+ 9{Q_{m}^2} |\Lambda|-
162 {Q_{m}^2} q
|\Lambda|^2}{27 |\Lambda|^3} \,, \quad {\cal P} \equiv -
\frac{1+3{Q_{m}^2} |\Lambda|}{3|\Lambda|^2}\,.
\label{cardano3}
\end{equation}
The number of real roots is known to be controlled by the
discriminant $\Delta = \frac{{\cal P}^3}{27} + \frac{{\cal
Q}^2}{4}$. There are three real roots, when $\Delta \leq 0$,
two roots coincide when the discriminant vanishes
$\Delta=0$, and
there is one real root when $\Delta>0$.

Let us consider one interesting example
exactly solvable. We put ${\cal Q}=0$, i.e.,
\begin{equation}
|\Lambda|=\frac{1}{36q}\left(1+ \sqrt{1+\frac{16q}{{Q_{m}^2}}}
\right)\,.\label{D2}
\end{equation}
Then $\Delta=\frac{{\cal P}^3}{27}<0$, and we obtain from
(\ref{cardano1})--(\ref{cardano3}) three real roots
\begin{gather}
X_{1} = -\frac{1}{3|\Lambda|} +
\sqrt{\frac{1+3{Q_{m}^2}
|\Lambda|}{3|\Lambda|^2}} \,, \quad X_{2}=-\frac{1}{3|\Lambda|}
\,, \nonumber\\
X_{3} = -\frac{1}{3|\Lambda|} -
\sqrt{\frac{1+3{Q_{m}^2} |\Lambda|}{3|\Lambda|^2}}\,.
\label{D3}
\end{gather}
The root $X_{1}$ is positive and thus the minimum of the
function $f(r,q,-|\Lambda|,Q_{m})$ gives
$r_{\rm c} = \sqrt{X_{1}}$, i.e.,
\begin{equation}
r_{\rm c}  =
\frac{1}{\sqrt{3|\Lambda|}}\sqrt{\sqrt{3+
{9}{Q_{m}^2}
|\Lambda|}-1} \,,\label{D4}
\end{equation}
which is the radius corresponding to
$M_{\rm c}$ in Eq.~(\ref{1reN00}) in this
particular case studied.

\subsection{The function $N(r)$ and the five distinct cases}

\vspace{3mm} \noindent
(i)  The case $M=0$

\noindent
When $M=0$, the metric function is
\begin{equation}
N(r)= 1+
\frac{r^2}{3}\,
\frac{3{Q_{m}^2}+ |\Lambda| r^4 }{r^4+{2Q_{m}^2} q}
\,.
\label{D19<0}
\end{equation}
Clearly, for $M=0$ and negative cosmological constant $N(r)\geq 1$
everywhere, and thus there are no horizons. The derivative
\begin{equation}
N^{\prime}(r) = \frac{2r}{(r^4{+}{2Q_{m}^2} q)^2} \left[\frac13
|\Lambda|r^8 {+} {Q_{m}^2} r^4 (2q |\Lambda|{-}1) {+}
{Q_{m}^4} q \right] \label{D5}
\end{equation}
is equal to zero at the center $r=0$, and
so this point is a local minimum for the metric function $N(r)$.
Now, there are still further features
depending on whether $6q|\Lambda|<1$, $6q|\Lambda|=1$,
or $6q|\Lambda|>1$.

\noindent
When $6q|\Lambda|<1$, there are two additional
extrema, the maximum at
\begin{equation}
r_{\rm max} {=}
\left\{\frac{3{Q_{m}^2}}{2|\Lambda|}
\left[(1{-}2q|\Lambda|)
{-}
\sqrt{(1{-}2q|\Lambda|)^2 {-}\frac83 q|\Lambda|} \right]
\right\}^{\frac14}\,,\label{D7}
\end{equation}
and the minimum at
\begin{equation}
r_{\rm min} {=}
\left\{\frac{3{Q_{m}^2}}{2|\Lambda|}
\left[(1{-}2q|\Lambda|) {+}
\sqrt{(1{-}2q|\Lambda|)^2 {-}\frac83 q|\Lambda|} \right]
\right\}^{\frac14}\,.\label{D9}
\end{equation}
In the limiting case $|\Lambda| \to 0$ the minimum drifts to
infinity, and the maximum coincides with the value $(2Q_{m}^2
q)^{\frac14}$, thus recovering the case with $\Lambda=0$.

\noindent
When
$6q|\Lambda|=1$, the two
radii above coincide with an inflection point, i.e.,
\begin{equation}
r_{\rm inflection}=r_{\rm max}=r_{\rm min}=(6{Q_{m}^2} q)^{\frac14}\,,
\label{Dinf}
\end{equation}
yielding
$N(r_{\rm inflection})= 1+\sqrt{\frac{{Q_{m}^2}}{6q}}$.

\begin{figure}[t]
\includegraphics[height=6cm]{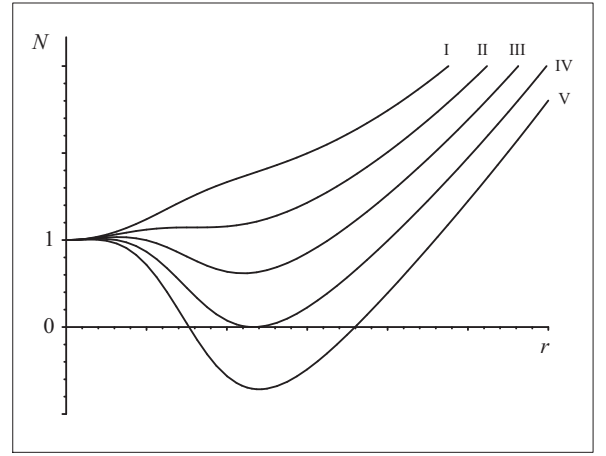}
\caption{$\Lambda<0$.
The plots of the metric function $N(r)$ for the nonminimal regular
models with $\Lambda<0$.
Note that $N(0)=1$ for all curves so that
the spacetimes are regular.  All the curves contain a small cavity
near the center, a repulsion barrier, and an AdS asymptotic behavior.
Curve I displays the behavior of
$N(r)$ when $M=0$.
Curve II, for $M=M_{\rm t}$
is also representative of the case $0<M<M_{\rm t}$.
There are no horizons, the objects are nonminimal stars.
Curve III, for
$M_{\rm t}<M<M_{\rm c}$ also represents nonminimal
stars, in addition it has a
flat well with a
repulsive barrier followed by a
Newtonian-type attraction zone.
Curve IV is the critical case $M=M_{\rm c}$ with a double horizon,
it describes an extremal black hole.
Curve  V
describes a black hole with $M_{\rm c}<M$, it has a
Cauchy horizon and an event horizon.
}
\label{fig4}
\end{figure}

\noindent
When
$6q|\Lambda|>1$ there are no additional extrema. This latter case,
$6q|\Lambda|>1$, is illustrated by curve I in Fig.~\ref{fig4}.

\vspace{3mm} \noindent
(ii) The case $0<M\leq M_{\rm t}$

\noindent
For $0<M\leq M_{\rm c}$
there appears
another specific mass
$M_{\rm t}$, with
${\rm t}$ denoting
triple, for which
$N^{\prime}(r_{\rm t})=0$ and $N^{\prime\prime}(r_{\rm
t})=0$. At this radius, the curve of the metric function
$N(r)$ has two
extrema coinciding with the inflection point.
For $M=M_{\rm t}$
the objects  correspond to regular stars
since there are no horizons, see curve II in
Fig.~\ref{fig4}.
We note that this curve is of the same type as the
curve for the case $M=0$ and $6q|\Lambda|=1$.
In the range  $0<M\leq M_{\rm t}$
there are no horizons in the spacetime also,
and so we also deal with regular stars,
of which  curve II is representative.

The plots of the metric function $N(r)$ for the nonminimal regular
models with $\Lambda=0$.  Note that $N(0)=1$ for all curves so that
the spacetimes are regular.  All the curves contain a small cavity
near the center, a repulsion barrier, and a flat asymptotic behavior.
Curve I displays the behavior of $N(r)$, when $M=0$.  Curve II is for
the case $0<M<M_{\rm c}$. There are no horizons in the spacetime, the
object is a regular star.  Curve III, for case $M=M_{\rm c}$, is the
critical case. There is a double horizon at the minimum of the metric
function $N(r)$, i.e., the Cauchy and event horizons coincide.  Curve
IV describes a black hole with mass exceeding the critical mass
$M>M_{\rm c}$, it contains a Cauchy horizon and an event horizon.
Curves II, III, and IV, which correspond to massive objects, have
zones of Newtonian-type attraction.
\vspace{3mm}
\noindent
(iii) The case $M_{\rm t}<M<M_{\rm c}$

\noindent
When $M_{\rm t}<M<M_{\rm c}$ there are no horizons in
the spacetime.  The plot of $N(r)$ in this case has a minimum at
$r_{\rm min}$, for which $N'(r_{\rm min})=0$, see
curve III in Fig.~\ref{fig4}.
Since at $r_{\rm min}$ one has
$N'(r_{\rm min})=0$, massive particles suffer no force
and can be at rest there. However, any small perturbation
puts the particle to oscillate near this equilibrium position.

\vspace{3mm} \noindent
(iv) The case $M=M_{\rm c}$

\noindent
When $M=M_{\rm c}$, we deal with an extremal regular black hole
in an asymptotically
AdS spacetime. The horizon is
a double horizon, see curve IV
in Fig.~\ref{fig4}. There are no specific points up to the second
order that distinguish the zone of Newtonian-type attraction and the
zone of cosmological attraction.

\vspace{3mm}
\noindent
(v) The case $M_{\rm c}<M$

\noindent
When $M_{\rm c}<M$ we deal with regular
black holes with a Cauchy horizon and an event horizon
in an asymptotically
AdS spacetime, see curve V in
Fig.~\ref{fig4}.

\section{Conclusions}\label{disc}

The solution for the metric potential $N(r)$ that we have found,
namely $N= 1 +
\left(\frac{r^4}{r^4+{ 2Q_{m}^2} q}\right)
\left(-\frac{2M}{r}
+ \frac{{Q_{m}^2}}{r^2} -\frac{\Lambda}{3}r^2 \right)$,
presents a new exact spherically
symmetric static solution of a nonminimal $SU(2)$ Einstein--Yang-Mills
theory with a cosmological $\Lambda$ and a Wu-Yang ansatz.  The
expression for $N(r)$ shows explicitly a four-parameter family of
exact solutions. These parameters are the nonminimal parameter $q$,
the
cosmological constant $\Lambda$, the magnetic
charge $Q_{m}$, and
the mass $M$.

The most important feature of the family of exact solutions is that it
has solutions with horizons depending on the relative values of the
parameters. We would like to stress the following details in this
context.

For $\Lambda>0$, depending on the values of the parameters, the black
hole solution can have three horizons, the Cauchy horizon, the event
horizon and the cosmological horizon.  When $M<M_{\rm c1}$, the first
critical mass value, there is only one horizon, the cosmological
horizon.  This means that a typical profile of the metric function
$N(r)$ contains a central small cavity, a repulsion barrier, a zone of
rest for matter near the point of minimum, a Newtonian-type attraction
zone, a zone of cosmological acceleration and a zone beyond the
cosmological horizon.  When $M=M_{\rm c1}$, there is a double extremal
horizon, formed by the Cauchy and event horizons together, in addition
to the cosmological one. When $M_{\rm c1}< M<M_{\rm c2}$ there are
then three separate horizons, the Cauchy, event and cosmological
horizons.  When $M=M_{\rm c2}$, there is a separate Cauchy horizon,
and there is another horizon coincidence, the event horizon and the
cosmological horizon come together.  In this case the black hole is a
cosmological extremal supermassive regular black hole as the whole
visible universe is swallowed by this supermassive object.  For $
M>M_{\rm c2}$ the Cauchy horizon is also a cosmological horizon, the
black hole is ultra massive. In such a universe there is only one
horizon, which is a cosmological one, together with a repulsion zone. In
this context this ultramassive black hole is a new type, in which
all three horizons coincide - the Cauchy, event, and cosmological
horizons.  There is also the case for which $M_{\rm c1}=M_{\rm c2}$
where the three horizons coincide.

For $\Lambda =0$, there is no cosmological horizon and depending on the
values of the parameters, the black hole solution can have two
horizons - the Cauchy horizon and the event horizon. When the mass is
below a critical mass $M_{\rm c}$, there are no horizons. For $M=M_{\rm
c}$ a double horizon appears, and when $M$ exceeds $M_{\rm c}$ there
are the Cauchy horizon and event horizons.

For $\Lambda <0$, there is also no cosmological horizon and depending
on the values of the parameters, the solution can have two
horizons - the Cauchy horizon and the event horizon. There is also a critical
mass $M_{\rm c}$, below which there are stars and above
which there are regular black holes with two horizons. The critical
case is a black hole with a double extremal horizon.

Geodesics of massive and massless test particles in these regular
black hole geometries display a great variety of cases.  Test
particles can be in a state of rest on the bottom of the well, or
oscillate around this minimum, other particles can overcome the
repulsion barrier and be lodged in the central small cavity, and
several other trajectories are possible.  However, particles cannot
leave the well into the exterior since there is a black hole
horizon, i.e., there is a one-way membrane.

\appendix

\acknowledgments

ABB and AEZ thank financial support from the Program of Competitive
Growth of Kazan Federal University (KFU)
Project No.~0615/06.15.02302.034 and from the Russian
Foundation for Basic Research Grant (RFBR) No.~14-02-00598.  ABB
acknowledges financial support provided under the European Union's
Framework Program 7
(FP7) European Research Council
(ERC) Starting Grant ``The dynamics of black holes:~testing the
limits of Einstein's theory" grant agreement No.~DyBHo-256667.
JPSL thanks Funda\c c\~ao para a Ci\^encia e Tecnologia
(FCT) - Portugal for financial support through Project
No.~PEst-OE/FIS/UI0099/2014.

\end{document}